\documentclass[conference,letterpaper]{IEEEtran}

\usepackage[utf8]{inputenc} 
\usepackage[T1]{fontenc}
\usepackage{url}
\usepackage{ifthen}
\usepackage{cite}
\usepackage[cmex10]{amsmath} 

\usepackage[dvipsnames]{xcolor}
\usepackage{bbm}

\usepackage{mdframed}

\newenvironment{weirdFrame}[1]
  {\mdfsetup{
    frametitle={\colorbox{white}{\space#1\space}},
    innertopmargin=0pt,
    skipabove=\topskip, skipbelow=\topskip,
    nobreak=true,
    frametitleaboveskip=-\ht\strutbox,
    frametitlealignment=\center
    }
  \begin{mdframed}%
  }
  {\end{mdframed}}

\newtheorem{theorem}{Theorem}

\newtheorem{definition}[theorem]{Definition}
\newtheorem{construction}[theorem]{Construction}
\newtheorem{assumption}[theorem]{Assumption}

\usepackage{lipsum}

\IEEEoverridecommandlockouts 

\makeatletter
\def\thanks#1{\protected@xdef\@thanks{\@thanks
        \protect\footnotetext{#1}}}
\makeatother

\makeatletter
\renewcommand\footnoterule{\relax\kern-5pt
\hrule
\kern4.6pt}
\makeatother

\begin{document}

\newcommand{\justin}[1]{\textcolor{blue}{#1}}

\interdisplaylinepenalty=2500 
\title{Amortized Locally Decodable Codes}

\author{%
 \IEEEauthorblockN{Jeremiah Blocki and Justin Zhang}
\IEEEauthorblockA{Department of Computer Science\\
                   Purdue University\\
                   West Lafayette, IN\\
                   Emails: \texttt{\{jblocki,zhan3554\}@purdue.edu}}}
\thanks{Supported in part by NSF CAREER Award CNS-2047272.}


\newcommand{\LCC}{\msf{LCC}}

\maketitle

\newcommand{\todo}[1]{\textcolor{red}{\underline{Todo}: #1}}
\newcommand{\jadd}[1]{\textcolor{PineGreen} {\underline{added}: #1}}
\newcommand{\jrem}[1]{\textcolor{purple} {\underline{Justin removed:} #1}}

\newcommand{\new}[1]{\textcolor{MidnightBlue}{#1}}
\newcommand{\jeremiah}[1]{\textcolor{red}{#1}} 

\newcommand{\gdGame}{\texttt{GD-Game}}
\newcommand{\prf}{\msf{prf}}
\newcommand{\aldc}{\msf{aLDC}}
\newcommand{\paldc}{\msf{paLDC}}
\newcommand{\paLDC}{\paldc}
\newcommand{\paldcgame}{\texttt{paLDC-Sec-Game}}
\newcommand{\Range}{\text{Range}}
\newcommand{\inv}[1]{#1^{-1}}
\newcommand{\RSE}{\msf{RSE}}
\newcommand{\RSEGame}{\texttt{RSE-Game}}
\newcommand{\ADPGame}{\texttt{ADP-Game}}

\newcommand{\Puz}{\msf{Puz}}
\newcommand{\Sol}{\msf{Sol}}

\newcommand{\code}[1]{\ensuremath{\calC_{\msf{#1}}}}
\newcommand{\encode}[1]{\ensuremath{\Enc_{\msf{#1}}}}
\newcommand{\decode}[1]{\ensuremath{\Dec_{\msf{#1}}}}
\newcommand{\en}[1]{\ensuremath{n_{\msf{#1}}}}
\newcommand{\kay}[1]{\ensuremath{k_{\msf{#1}}}}

\newcommand{\LDCStar}{\ensuremath{\msf{LDC}^*}}
\newcommand{\calCStar}{\code *}
\newcommand{\EncStar}{\encode *}
\newcommand{\DecStar}{\decode *}
\newcommand{\nStar}{\en *}
\newcommand{\kStar}{\kay *}
\newcommand{\YStar}{Y_{*}}
\newcommand{\ellStar}{\ell_*}
\newcommand{\deltaStar}{\delta_*}
\newcommand{\epsStar}{\eps_*}

\newcommand{\calCP}{\ensuremath{\calC_{\msf P}}}

\newcommand{\EncP}{\encode P}
\newcommand{\DecP}{\decode P}
\newcommand{\nP}{\en P}
\newcommand{\kP}{\kay P}
\newcommand{\YP}{Y_{\msf P}}
\newcommand{\kappaP}{\kappa_{\msf{P}}}
\newcommand{\alphaP}{\alpha_{\msf{P}}}
\newcommand{\deltaP}{\delta_{\msf{P}}}
\newcommand{\epsP}{\eps_{\msf{P}}}

\newcommand{\epsPuz}{\eps_{\msf{Puz}}}


\newcommand{\eps}{\varepsilon}
\newcommand{\inprod}[1]{\left\langle #1 \right\rangle}
\newcommand{\m}[1]{\begin{bmatrix}#1\end{bmatrix}}

\newcommand{\msf}[1]{\ensuremath{\mathsf{#1}}}
\newcommand{\mc}{\mathcal}

\newcommand{\calA}{\mathcal{A}}
\newcommand{\calB}{\mathcal{B}}
\newcommand{\calC}{\mathcal{C}}
\newcommand{\calD}{\mathcal{D}}
\newcommand{\calE}{\mathcal{E}}
\newcommand{\calF}{\mathcal{F}}
\newcommand{\calG}{\mathcal{G}}
\newcommand{\calH}{\mathcal{H}}
\newcommand{\calI}{\mathcal{I}}
\newcommand{\calJ}{\mathcal{J}}
\newcommand{\calK}{\mathcal{K}}
\newcommand{\calL}{\mathcal{L}}
\newcommand{\calM}{\mathcal{M}}
\newcommand{\calN}{\mathcal{N}}
\newcommand{\calO}{\mathcal{O}}
\newcommand{\calP}{\mathcal{P}}
\newcommand{\calQ}{\mathcal{Q}}
\newcommand{\calR}{\mathcal{R}}
\newcommand{\calS}{\mathcal{S}}
\newcommand{\calT}{\mathcal{T}}
\newcommand{\calU}{\mathcal{U}}
\newcommand{\calV}{\mathcal{V}}
\newcommand{\calW}{\mathcal{W}}
\newcommand{\calX}{\mathcal{X}}
\newcommand{\calY}{\mathcal{Y}}
\newcommand{\calZ}{\mathcal{Z}}

\newcommand{\R}{\mathbbm R}
\newcommand{\C}{\mathbbm C}
\newcommand{\N}{\mathbbm N}
\newcommand{\Z}{\mathbbm Z}
\newcommand{\F}{\mathbbm F}
\newcommand{\Q}{\mathbbm Q}

\newcommand{\bone}{\boldsymbol{1}}
\newcommand{\bbeta}{\boldsymbol{\beta}}
\newcommand{\bdelta}{\boldsymbol{\delta}}
\newcommand{\bepsilon}{\boldsymbol{\epsilon}}
\newcommand{\blambda}{\boldsymbol{\lambda}}
\newcommand{\bomega}{\boldsymbol{\omega}}
\newcommand{\bpi}{\boldsymbol{\pi}}
\newcommand{\bphi}{\boldsymbol{\phi}}
\newcommand{\bvphi}{\boldsymbol{\varphi}}
\newcommand{\bpsi}{\boldsymbol{\psi}}
\newcommand{\bsigma}{\boldsymbol{\sigma}}
\newcommand{\btheta}{\boldsymbol{\theta}}
\newcommand{\btau}{\boldsymbol{\tau}}
\newcommand{\ba}{\boldsymbol{a}}
\newcommand{\bb}{\boldsymbol{b}}
\newcommand{\bc}{\boldsymbol{c}}
\newcommand{\bd}{\boldsymbol{d}}
\newcommand{\be}{\boldsymbol{e}}
\newcommand{\boldf}{\boldsymbol{f}}
\newcommand{\bg}{\boldsymbol{g}}
\newcommand{\bh}{\boldsymbol{h}}
\newcommand{\bi}{\boldsymbol{i}}
\newcommand{\bj}{\boldsymbol{j}}
\newcommand{\bk}{\boldsymbol{k}}
\newcommand{\bell}{\boldsymbol{\ell}}
\newcommand{\bm}{\boldsymbol{m}}
\newcommand{\bn}{\boldsymbol{n}}
\newcommand{\bo}{\boldsymbol{o}}
\newcommand{\bp}{\boldsymbol{p}}
\newcommand{\bq}{\boldsymbol{q}}
\newcommand{\br}{\boldsymbol{r}}
\newcommand{\bs}{\boldsymbol{s}}
\newcommand{\bt}{\boldsymbol{t}}
\newcommand{\bu}{\boldsymbol{u}}
\newcommand{\bv}{\boldsymbol{v}}
\newcommand{\bw}{\boldsymbol{w}}
\newcommand{\bx}{{\boldsymbol{x}}}
\newcommand{\by}{\boldsymbol{y}}
\newcommand{\bz}{\boldsymbol{z}}
\newcommand{\bA}{\boldsymbol{A}}
\newcommand{\bB}{\boldsymbol{B}}
\newcommand{\bC}{\boldsymbol{C}}
\newcommand{\bD}{\boldsymbol{D}}
\newcommand{\bE}{\boldsymbol{E}}
\newcommand{\bF}{\boldsymbol{F}}
\newcommand{\bG}{\boldsymbol{G}}
\newcommand{\bH}{\boldsymbol{H}}
\newcommand{\bI}{\boldsymbol{I}}
\newcommand{\bJ}{\boldsymbol{J}}
\newcommand{\bK}{\boldsymbol{K}}
\newcommand{\bL}{\boldsymbol{L}}
\newcommand{\bM}{\boldsymbol{M}}
\newcommand{\bN}{\boldsymbol{N}}
\newcommand{\bP}{\boldsymbol{P}}
\newcommand{\bQ}{\boldsymbol{Q}}
\newcommand{\bR}{\boldsymbol{R}}
\newcommand{\bS}{\boldsymbol{S}}
\newcommand{\bT}{\boldsymbol{T}}
\newcommand{\bU}{\boldsymbol{U}}
\newcommand{\bV}{\boldsymbol{V}}
\newcommand{\bW}{\boldsymbol{W}}
\newcommand{\bX}{\boldsymbol{X}}
\newcommand{\bY}{\boldsymbol{Y}}
\newcommand{\bZ}{\boldsymbol{Z}}

\newcommand{\ignore}[1]{}

\newcommand{\unif}{\overset{{\scriptscriptstyle\$}}{\leftarrow}}

\newcommand{\sargram}{\msf{SArg}-\msf{RAM}}
\newcommand{\poly}{\msf{poly}}
\newcommand{\polylog}{\msf{polylog}}
\newcommand{\negl}{\msf{negl}}

\makeatletter
\newcommand{\subalign}[1]{%
  \vcenter{%
    \Let@ \restore@math@cr \default@tag
    \baselineskip\fontdimen10 \scriptfont\tw@
    \advance\baselineskip\fontdimen12 \scriptfont\tw@
    \lineskip\thr@@\fontdimen8 \scriptfont\thr@@
    \lineskiplimit\lineskip
    \ialign{\hfil$\m@th\scriptstyle##$&$\m@th\scriptstyle{}##$\hfil\crcr
      #1\crcr
    }%
  }%
}
\makeatother

\newcommand{\LDC}{\ensuremath{\msf{LDC}}}
\newcommand{\aLDC}{\ensuremath{\msf{aLDC}}}
\newcommand{\pLDC}{\ensuremath{\msf{pLDC}}}
\newcommand{\apLDC}{\ensuremath{\msf{paLDC}}}
\newcommand{\Gen}{\ensuremath{\msf{Gen}}}
\newcommand{\Enc}{\ensuremath{\msf{Enc}}}
\newcommand{\Dec}{\ensuremath{\msf{Dec}}}
\newcommand{\sk}{\ensuremath{\msf{sk}}}
\newcommand{\pk}{\ensuremath{\msf{pk}}}
\newcommand{\adv}{\ensuremath{\calA}}
\begin{abstract}
Locally Decodable Codes (LDCs) are error correcting codes that admit efficient decoding of individual message symbols without decoding the entire message. 
Unfortunately, known LDC constructions offer a sub-optimal trade-off between rate, error tolerance and locality, the number of queries that the decoder must make to the received codeword $\tilde {\by}$ to recovered a particular symbol from the original message $\bx$, even in relaxed settings where the encoder/decoder share randomness or where the channel is resource bounded. 
We initiate the study of Amortized Locally Decodable Codes where the local decoder wants to recover multiple symbols of the original message $\bx$ and the total number of queries to the received codeword $\tilde{\by}$ can be amortized by the total number of message symbols recovered. 
We demonstrate that amortization allows us to overcome prior barriers and impossibility results. 
We first demonstrate that the Hadamard code achieves amortized locality below $2$ --- a result that is known to be impossible without amortization. 
Second, we study amortized locally decodable codes in cryptographic settings where the sender and receiver share a secret key or where the channel is resource-bounded and where the decoder wants to recover a consecutive subset of message symbols $[L,R]$. 
In these settings we show that it is possible to achieve a trifecta: constant rate, error tolerance and constant amortized locality. 
\end{abstract}

\section{Introduction}
Locally Decodable Codes ($\LDC$s) are error correcting codes that admit fast single-symbol decodability after making a small number of queries to the received (possibly corrupted) codeword $\tilde{\by}$. In particular, an $(n,k)$-code over an alphabet $\Sigma$ is an $(\ell,\delta,\eps)$-\LDC\ if there exists a pair of sender/receiver algorithms $\Enc: \Sigma^k \rightarrow \Sigma^n$ encoding messages of length $k$ to codewords of length $n$, and $\Dec^{\tilde{\by}}: [k] \rightarrow \Sigma$ decoding any requested single message index $i \in [k]$ where $[k] := \{1,2,\dots,k\}$.  We require that for all messages $\bx$ and received word $\tilde {\by},$ the decoder makes at most $\ell$ queries to $\tilde {\by}$ and if the hamming distance $d(\Enc(\bx),\tilde {\by}) \leq \delta n$ i.e the error caused by an adversarial channel is at most $\delta n$, we require that the decoder is correct with probability at least $1 - \eps$ i.e $\Pr[\Dec^{\tilde {\by}}(i) = \bx_i] \geq 1 -\eps.$  The main parameters of interest in \LDC s are the {\em rate} $R := k / n$,  {\em locality} $\ell$, and {\em error-rate tolerance} $\delta$.
For instance, $\LDC$ s can be used to store files, where we want the rate to be as close to $1$ as possible to limit storage overhead, the error-tolerance to be high for fault tolerance, and locality to be low for ultra-efficient recovery of any portion of the file.

The trade-offs between rate, locality, and error tolerance within \LDC s for (worst-case) classical errors have been extensively studied yet achievable parameters remain sub-optimal and undesirable.
Ideally, we would like an \LDC\ which achieves constant rate, constant locality, and constant error-rate tolerance simultaneously.
However, any \LDC\ with constant locality $\ell \geq 2$ or constant error tolerance $\delta > 0$ must have sub-constant rate \cite{STOC:KatTre00}, where the best known constructions (e.g Hadamard and matching vector codes) have super-polynomial codeword length \cite{LDCSURVEY}.
In particular, for locality $\ell = 2$, any construction must have exponential rate \cite{ STOC:KerdWo03,CCC:GKST06,FOCS:BenRegdWo08}.
Katz and Trevisan show further that there do not exist \LDC\ for $\ell < 2$ even when the rate is allowed to be exponential\cite{STOC:KatTre00}.
It is easy to verify that their result further extends to settings where the error-rate is $\delta = o(1)$ e.g $\delta = O(1/n^{.99}).$
In other words, a local decoder must read {\em at least twice} as much information as requested in the setting of worst-case errors.

Various relaxations have been introduced to deal with these undesirable trade-offs for classical \LDC s.
For example, Ben-Sasson et al. introduced the notion of a relaxed \LDC s\cite{STOC:BGHSV04}, allowing the decoder to reject (output $\bot$) instead of outputting a codeword symbol whenever it detects error.
This was further expanded by Gur et al. to study locally {\em correctable} codes where the decoder returns symbols of the codeword instead of the message \cite{ITCS:GurRamRot18}.
Another line of work studies \LDC s that allow a sender and receiver to share a secret key that is unknown to the computationally-bound channel \cite{EPRINT:OstPanSah07,C:HemOst08,public_key_ldc_w_short_keys}.
Yet another line of work considers \LDC s in settings where the channel is resource bounded (e.g it cannot evaluate circuits beyond a particular size/depth) \cite{ITC:BloKulZho20}.

However, even with these relaxations, the achievable trade-offs are still sub-optimal.
For relaxed locally decodable/correctable codes respectively, Ben-Sasson et al. and Gur et al. are able to achieve \LDC/\LCC\ constructions with constant locality and constant error-tolerance codes, but with sub-optimal codeword length $n = O\left(k^{1 + O(1)/\sqrt \ell}\right)$ \cite{SODA:ChiGurShi20, STOC:BGHSV04}.
Gur and Lachish also prove that any relaxed \LDC\ must have codeword length $n = \Omega\left(k^{1 + c}\right)$, where $c = 1 / O\left(\ell^2\right)$ \cite{SODA:GurLac20}, ruling out any possibility of having constant rate, constant locality, and constant error-tolerance.
In fact, any relaxed \LDC\ with constant error tolerance, perfect completeness, and locality $\ell = 2$ must have exponential rate \cite{CCC:BBCGLZZ23}. 
{Notably, constructions for relaxed \LDC s with constant error-rate tolerance, constant rate, and polylogarithmic locality $\ell = O(\polylog(k))$ exist in the shared-cryptographic-key \cite{EPRINT:OstPanSah07,C:HemOst08,public_key_ldc_w_short_keys}, resource-bounded-channel settings \cite{SCN:AmeBloBlo22,ITC:BloKulZho20}, and more recently, in the information theoretic setting with no additional assumptions \cite{STOC:KumMon24,CCC:CohYan24}.
However, there are no constant rate, constant error-rate tolerance, constant locality relaxed \LDC \ constructions \cite{SODA:GurLac20}. }

Traditionally, in \LDC\ literature, the decoder is tasked with recovering the symbol or bit at a single index.
However, most practical applications desire the recovery of a much larger portion of the message.
For example, the local decoder may want to recover the symbol $\bx_i$ for {\em every} index $i \in Q$ for some set $Q \subseteq [k]$.
The natural and naive way to accomplish this task would be to run a local decoder $|Q|$ times separately for each $i \in Q$, but the total query complexity will be $\ell |Q|$. 
In this paper, we ask the following natural question: when $|Q| > 1$, is it possible to improve the total query complexity beyond that of the naive solution by designing a decoder that attempts to decode all requested symbols in one run, {\em amortizing} the number of queries?

\subsection{Our Contributions}
We initiate the study of {\em amortized locally decodable codes} which seek to reduce query complexity by amortizing the local decoding process. 
Given a set $Q \subseteq [k]$, the local decoding algorithm $\Dec^{\tilde{\bY}}(Q)$ should output $\{\bx_i\}_{i \in Q}$ where the amortized query complexity $\alpha$ is given by the total number of queries made by the decoder $\ell$ divided by the total number of message symbols recovered i.e., $\alpha = \ell/|Q|$.  

We first show that the Hadamard code \cite{LDCSURVEY} can achieve {\em amortized} locality $\alpha < 2$ when $|Q| > 2$. 
In fact, if the the error-rate is $\delta=o(1)$ then the amortized locality approaches $1$ when $|Q|$ is sufficiently large. 
This stands in stark contrast to the impossibility results of Katz and Trevisan who proved that, without amortization, any \LDC\ must have $\ell \geq 2$ even if $\delta=o(1)$.  

Second, we study amortized locally decodable codes in cryptographic settings where the sender and receiver share a secret key. 
We show that when the decoder wants to recover a {\em consecutive} subset of bits $[L,R] \subset [k]$ that it is possible to achieve {\em constant rate, constant error tolerance and constant amortized locality} as long as $R-L \geq \kappa$  is sufficiently large e.g., $\kappa = \omega(\log k)$. To the best of our knowledge this is the first construction which achieves all three goals simultaneously, even in settings where the sender/receiver share a secret key. 

Finally, we can apply the framework of \cite{SCN:AmeBloBlo22,ITC:BloKulZho20,ISIT:BloBlo21} to remove the assumption that the sender and receiver share a secret key as long as the channel is resource bounded and is unable to solve cryptographic puzzles. As Blocki et al. \cite{ITC:BloKulZho20} argued, resource bounded channels can plausibly capture any error pattern that arises naturally, that is, real-world channels are resource bounded. For example, suppose that $A$ denotes a randomized algorithm that models the error pattern of our channel. If the channel has small latency then we can reasonably assume that the algorithm $A$ must have low-depth --- there may be many computational steps if the algorithm $A$ is parallel, but the depth of the computation is bounded. This means that a low-latency channel would be incapable of solving  time-lock puzzle \cite{rivest1996time} --- cryptographic puzzles that are solvable in $t$ sequential computation steps, but cannot be solved in $o(t)$ time by any parallel algorithm running in polynomial time. One can also design cryptographic puzzles that are space-hard meaning that they cannot be solved by any probabilistic polynomial time algorithm using space $o(s)$, but can be solved easily using space $s$. Additionally, other categories of resource-bounded cryptographic puzzles exist, such as memory-hard \cite{SCN:AmeBloBlo22}, space-hard puzzles, which impose constraints on time-space complexity and space complexity respectively. Ameri et al. \cite{SCN:AmeBloBlo22} showed how to use cryptographic puzzles and secret key \LDC s to construct resource-bounded \LDC s with constant rate, constant error tolerance, but their locality is $O(\text{polylog} (k))$. We demonstrate that this construction achieves amortized locality $O(1)$ if we use our amortizable secret key \LDC.  

\subsection{Relationship to Batch Codes} (Smooth) Locally Decodable Codes have been used to construct Private Information Retrieval (PIR) protocols in the information theoretic setting with multiple servers. Ishai et al. \cite{STOC:IKOS04} introduced the notion of batch codes as a tool to minimize the maximum (or amortized) workload on any individual PIR server. Intuitively, a $(k,N,\kappa, m)$ batch code encodes a string $\bx \in \Sigma^k$ into a $m$ tuple $\by_1,\ldots, \by_m \in \Sigma^*$ (buckets) of total length $N$ such that for any set of indices $i_1,\ldots, i_\kappa \in [k]$ the entries $x_{i_1},\ldots, x_{i_\kappa}$ can be decoded by reading {\em at most one symbol} from each bucket. However, it seems unlikely that batch codes could be used to constructed amortized \LDC s as the definition is quite different e.g., there is no notion of errors. On the flip side it is currently unclear whether or not amortized \LDC s can usefully applied to construct PIR protocols.   

\section{Amortized Locality}\label{sect:aLDC}
We provide the first formalization of amortized locally decodable codes. 
We say two strings $g$ and $h$ of the same length are $\delta$-close if $g$ has hamming distance at most $\delta |g|$ from $h$.
    \begin{definition}
    A $(n,k)$-code $\calC$ is a $(\alpha, \kappa, \delta, \epsilon)$-amortizeable LDC (\aLDC) if there exists an algorithm $\Dec$ such that for every $x \in \Sigma^k,$ $\tilde y \in \Sigma^n$ such that $\tilde{\by}$ is $\delta$-close to $\calC(x),$ and every subset $Q \subseteq [k]$ with $|Q| \geq \kappa$ we have 
    $\Pr\lbrack\msf \Dec^{\tilde y}(Q) = \{\bx_i : i \in Q\}\rbrack \geq 1 - \eps$
and $\Dec^{\tilde y}$ makes at most $\alpha|Q|$ queries to $\tilde y$. 
\end{definition}
An $(\alpha,\kappa,\delta,\eps)$-\aLDC\ permits that the decoder make up to $|Q| \alpha$ total queries when attempting to decode the target symbols in the set $Q$. The amortized number of queries per symbol is just $\alpha$, but because the decoder may make up to $|Q| \alpha$ queries in total, it may be possible to circumvent classical barriers and impossibility results. 

As a first motivation for \aLDC s we first consider an impossibility result of Katz and Trevisan \cite{STOC:KatTre00} who proved that any LDC must have locality $\ell \geq 2$. In particular, they proved that for any $(1,\delta,\epsilon)$-\LDC\ we have $k \leq \frac{\log |\Sigma|}{\delta(1- H(1/2 + \eps))}$ where $H(\cdot)$ is the entropy function in base $|\Sigma|$. Even if we set the error-tolerance to $\delta = 1/\sqrt{k}$, so that $\delta=o(1)$, we still have the constraint that $\sqrt{k} \leq \frac{\log |\Sigma|}{1-H(1/2+\epsilon)}$. Thus, it is impossible to construct a $(1,\delta,\epsilon)$-\LDC\ which supports arbitrarily long messages in $\Sigma^k$. It follows that any $\LDC$ construction that supports long messages must have locality $\ell \geq 2$. We show that it is possible to break this barrier by amortizing the decoding costs across multiple queries and achieve amortized locality $1+ O(\delta/\epsilon)$. Note that if $\epsilon$ is a constant and $\delta=o(1)$, the amortized locality is $1+o(1)$, that is, the amortized locality approaches $1$. In fact, we show that the Hadamard code already achieves these properties \cite{hadamard} --- see Theorem \ref{thm:hadamard}.

The Hadamard code encodes binary messages $\bx \in \{0,1\}^k$ of length $k$ to binary codewords $\by$ of length $2^k$, where each codeword bit corresponds to a XOR of message bit subsets, $\by_S := \oplus_{i \in S} \bx_i, \forall S \subseteq [k]$. More precisely, $\Enc(\bx) = \langle \by_S \rangle_{S\subseteq [k]} \in \{0,1\}^{2^k}$ where $\by_S := \oplus_{i \in S} \bx_i$. We show that by extending a simple idea from a traditional single bit local decoder, we can have amortize beyond what is possible for traditional locality. 

A simple decoder achieving $2$-locality for Hadamard codes decodes any message bit index $i$ by selecting a random subset $S \subseteq [k]$, computing the subset $S_i := S \Delta \{i\}$ ($\Delta$ denotes symmetric-difference), querying the received codeword $\tilde{\by}$ at the indices corresponding to $S$ and $S_i$ to obtaining $\tilde{\by}_S$ and $\tilde{\by}_{S_i}$ and then outputting $\tilde{\by}_S \oplus \tilde{\by}_{S_i}$. If the error-rate is set to $\delta,$ then by a union bound, with probability at least $1 - 2 \delta$ we have $\tilde{\by}_S = \by_S$ and $\tilde{\by}_{S_i} = \by_{S_i}$ i.e., both queried bits are correct. If both queried bits are correct, then the decoder will succeed as $\tilde{\by}_S \oplus \tilde{\by}_{S_i} = {\by}_S \oplus \by_{S_i}= \bx_i$. 

If we want to recover multiple message bits, then we can instead pick a random set $S$ and then query to obtain $\tilde{\by}_S$ and $\tilde{\by}_{S_j}$ for all $j \in Q$. The total number of queries will be $|Q|+1$ so the amortized locality is just $1+1/|Q|$. By union bounds we will have $\tilde{\by}_S = \by_S$ and $\tilde{\by}_{S_j} = \by_{S_j}$ for all $j \in Q$ with probability at least $1-\delta(|Q|+1)$. These observations lead to Theorem \ref{thm:hadamard}. 
\begin{theorem} \label{thm:hadamard}
    For any $k,\delta,\kappa > 0$, the Hadamard code is a $(2^k - 1,k)$-code that is also a $\left(\frac{\kappa + 1}{\kappa}, \kappa, \delta, \eps \right)$-\aLDC, where $\eps \geq (\kappa + 1) \delta.$
\end{theorem}
For example, if $\eps \leq \frac{1}{3}$ and $\delta \leq \frac{1}{9}$ then we have $\left(\frac
3 2, 2, \delta,\eps\right)$-\aLDC. In fact, if $\delta = o(1),$ then we have an \aLDC\ with amortized locality $\alpha \rightarrow 1$ as we can set $\kappa + 1 = \epsilon/\delta$ so that $\kappa \rightarrow \infty$. This is in stark contrast to the result of Katz and Trevisan, which state that (without amortization) no \LDC with locality $\ell < 2$ exists even when $\delta = o(1).$ \\ 
{\bf Optimality:} As a trivial observation observe that we cannot hope to achieve amortized locality $\alpha < 1$ when $\epsilon < \frac{1}{2}$. In particular, any \aLDC\ decoder that recovers a $\kappa$ bits of an arbitrary binary message $\bx$ with probability $1-\epsilon > \frac{1}{2}$ {\em must} make {\em at least} $\kappa$ queries to the codeword --- otherwise it would be possible to use the \aLDC\ to compress random $\kappa$ bit strings. 

\section{Private Locally Decodable Codes}\label{sect:paLDC}
In the previous section we saw how amortization allowed us to push past the locality $\ell=2$ barrier and achieve amortized locality $\alpha < 2$ with constant error-tolerance, The primary downside to the Hadamard construction is that the rate $R$ is exponential. Ideally, we want a construction with constant rate, constant error tolerance and constant amortized locality. It remains an open question whether or not this goal is achievable. As an initial step we show that the goal is achievable in relaxed settings where the sender and receiver share randomness or where the channel is computationally bounded. In this section we will also make the natural assumption that the decoder wants to recover a consecutive portion of the original message i.e., $Q = [L,R] = \{L, L+1,\ldots, R\} \subseteq [k]$. 

In settings where the sender and receiver share randomness (e.g., cryptographic keys) or where the channel is resource-bounded we can slightly relax the correctness condition for an \aLDC. Recall that we previously required the decoder $\Dec^{\tilde{\by}}$ succeed with probability at least $1-\epsilon$ for any corrupted codeword $\tilde{\by}$ that is sufficiently close to the original codeword $\by$. In relaxed versions of the definition, it is acceptable if there exists corrupted codewords $\tilde{\by}$ that fool the decoder and are close to the original codeword, as long as it is computationally infeasible for an adversarial, but resource-bounded channel to find such a corruption with high probability. This motivates definition \ref{def:paLDC}. 
   \begin{definition}\label{def:paLDC}
    Let $\lambda$ be the security parameter. A triple of probabilistic polynomial time algorithms $(\Gen,\Enc,\Dec)$ is a private $(\alpha, \kappa, \delta, \epsilon,q)$-amortizeable LDC (\paldc) if 
    \begin{itemize}
        \item for all keys $\sk \in \Range(\Gen(1^\lambda))$ the pair $(\Enc_{\sk},\Dec_{\sk})$ is an $(n,k)$-code, and
        \item for all probabilistic polynomial time algorithms $\calA$ there is a negligible function $\mu$ such that
        $\Pr[\paldcgame(\calA,\lambda,\delta,\kappa,q) = 1] \leq \mu(\lambda),$
        where the probability is taken over the randomness of $\calA, \Gen,$ and $\paldcgame$. 
        The experiment \paldcgame\ is defined as follows:
    \end{itemize}
     \begin{weirdFrame}%
        {\paldcgame$(\calA,\lambda,\delta,\kappa,q)$}
        The challenger generates secret key $\msf{sk} \leftarrow \Gen(1^\lambda)$. 
        For $q$ rounds, on iteration $h$, the challenger and adversary $\calA$ interact as follows:
        \begin{enumerate}
            \item The adversary $\calA$ chooses a message $x^{(h)} \in \{0,1\}^k$ and sends it to the challenger.
            \item The challenger sends $y^{(h)} \leftarrow \Enc(\sk,x^{(h)})$ to the adversary.
            \item The adversary outputs $\Tilde{y}^{(h)} \in \{0,1\}^n$ with hamming distance at most $\delta n$ from $y^{(h)}$.
            \item If there exists $L^{(h)}, R^{(h)} \in [k]$ such that $R^{(h)} - L^{(h)} + 1 \geq \kappa$ and
            \[\hspace{-\leftmargin}\Pr\left[\Dec^{\Tilde{y}^{(h)}}_\sk(L^{(h)},R^{(h)}) \neq x^{(h)}_L\cdots x^{(h)}_R\right] > \eps(\lambda)\] such that $\Dec^{\Tilde{y}^{(h)}}_\sk(.)$ makes at most $(R - L + 1) \alpha$ queries to $\tilde y$, then this experiment outputs $1$.
        \end{enumerate}
        If the experiment did not output $1$ on any iteration $h$, then output $0.$
        \end{weirdFrame}
        \end{definition}
   
\subsection{One-Time \paLDC }\label{subsect:oneTimePaLDC}
Our first \paLDC\ construction will be based on the private-key construction of Ostrovsky et al. \cite{ICALP:OstPanSah07}. 
The secret key in our scheme will be a random permutation $\pi$ and a one-time pad $\bR$. 
To encode the message $\bx,$ first split it into $B$ equal-sized blocks of size $a = \omega(\log \lambda)$, where $\bx = \bw_1 \circ \dots \bw_B$ ($\circ$ is the concatenation function).
Encode each block $\bw_j$ as $\bw'_j = \calC(\bw_j)$, where $\calC$ is a code with a constant rate $R$ and constant error-tolerance $\delta$. 
Form the encoded message as $\by' = \bw'_1 \circ\dots\circ \bw'_B,$ where $|\by| / |\bx| = 1 / R$. 
Note that we cannot just output $\by'$ without assuming sub-constant error-tolerance because otherwise, an adversarial channel can just choose to corrupt an entire block $\bw'_j.$
To remedy this, we apply our secret random permutation $\pi$ and the one-time pad $\bR$ to output $\by = \pi(\by' \oplus \bR)$.
Since the channel is computationally bounded, the errors it causes is effectively random. 
If the overall error tolerance is a constant dependent on $\delta$, then each block will have at most $\delta n$ errors with high probability (see \cite{ICALP:OstPanSah07} for details). 

Thus, our local decoder will simply recover its requested message symbols by recovering the corresponding message blocks.
That is, if block $\bw_j$ is requested, then we undo the permutation and one-time pad to obtain the encoded block $\bw'_j$ and subsequently decode it. 
More specifically, for each index $j_r$ of $\bw'_j$ in $\by'$, we obtain the corresponding index in $\by$ as $\pi(j_r).$
In summary, this code achieves constant rate, constant error-tolerance, and constant amortized locality with parameters summarized in the following theorem.
\begin{theorem}
    Suppose code $\calC$ has constant rate $R$ and constant error-tolerance. Then, the construction above is a $(2 / R, \omega(\log \lambda), O(1), O(1), 1)$-\paLDC.
\end{theorem}
The primary limitation to the above construction is that the security is only guaranteed after the encoder sends a {\em single} ($q=1$) message to the decoder. If the encoder has multiple messages to send, then they would need to use a separate permutation $\pi^i$ and a one-time pad $\bR^i$ in every round $i \leq q$.  
Generating and secretly sharing randomness for each message is costly and undesirable; instead, we propose an alternate where the secret key may be used polynomial many times.
The full analysis can be found in appendix \ref{app:oneTime}.
\subsection{Multi-Round \paLDC }\label{subsect:multPaLDC}
We present a polynomial round ($q = \poly(\lambda)$) \paLDC\ with constant rate, constant error rate tolerance, and constant amortized locality that matches the one-time construction without requiring multiple secret keys. Our primary technical ingredient is a special type of code that we call a {\em robust secret encryption (\RSE).} Intuitively, we want a code with the property that any computationally bounded adversary who does not have the secret key for the scheme cannot distinguish the encoding of a random message from a random string. This allows us to embed fresh randomness in such a way that the randomness is effectively hidden from an attacker who does not have the secret key. 

Formally, the definition of a \RSE\ is given in definition \ref{def:RSE}. Intuitively, the $\RSE$ game captures the property that a PPT attacker cannot distinguish the encoding of a random message from a truly random string even if the attacker is given many samples. Recent work \cite{C:ChrGun24} yields an efficient construction (constant rate/error tolerance) of \RSE\ from the Learning Parity with Noise (LPN) assumption --- a standard widely accepted assumption in the field of cryptography.

We also provide a construction based on the Gopppa code McEliece Cryptosystem, which can be found in appendix \ref{app:RSE}.
    \begin{definition}\label{def:RSE}
    A $(n,k,\delta)$-Robust Secret Encryption ($\RSE$) is a tuple of probabilistic polynomial time algorithms $(\Gen,\Enc,\Dec)$ such that:
    \begin{itemize}
        \item For all keys $\sk \in Range(\Gen(1^\lambda)),$ $(\Enc_\sk,\Dec_\sk)$ is a $(n,k)$ code that can tolerate $\delta n$ errors.
        \item For any probabilistic polynomial time algorithm $\calA$ playing the $\RSEGame$, $q \in \poly(\lambda),$ there exists a negligible function $\eps$ such that
        $\left| \Pr[\RSEGame(\calA,\lambda,q) = 1] - \frac{1}{2} \right| < \eps(\lambda),$
        where the $\RSEGame$ is defined as,
    \end{itemize}
    \vspace{-0.1in}
    \begin{weirdFrame}{$\RSEGame(\calA,\lambda,q)$}
    The challenger generates $\sk \leftarrow \Gen(1^\lambda)$ and $b \unif \{0,1\}$ then sends $\calA$ a sequence $\{R^i\}_{i \in [q]}$, where each $R^i$ is (identically and independently) generated as follows:
    \begin{itemize}
    \item if $b = 0$, $R^i \leftarrow \Enc_\sk(r^i)$ where $r^i \unif \F^k$,
    \item otherwise if $b = 1,$ $R^i \unif \F^n.$
    \end{itemize}
    $\calA$ outputs bit $b' \in \{0,1\}$, and if $b = b'$, the output of this experiment is $1$.
    Otherwise, the output is $0.$
    \end{weirdFrame}
\end{definition}
Lastly, we will need a common cryptographic tool known as the pseudorandom function (prf). 
Informally, a prf is a deterministic function $f$ that when instantiated with a secret key $\bk,$ is indistinguishable from a random function to a computationally-bound adversary. 
We use the prf to essentially generate a new one-time pad for each message sent, allowing us to invoke the indistinguishability of the \RSE.
    \begin{construction}\label{constr:RSE_paLDC} 
Let $\RSE.(\Gen,\Enc,\Dec)$ be an $(A,a,\delta)$-$\RSE$ with rate $R_\RSE$, and 
let $f : \{0,1\}^{\lambda} \times \{0,1\}^{b + \log B} \rightarrow \{0,1\}^{a}$ be a prf $f$, where $a = \poly(b)$.
\vspace{-0.13in}
\begin{weirdFrame}{$\Gen(1^\lambda) $}
    Output $\sk \leftarrow (\pi,\bk,\sk')$ where $\pi \unif S_{BA},$ $ \bk \unif \{0,1\}^\lambda$, and $\sk' \leftarrow \RSE.\Gen(1^\lambda)$.
\end{weirdFrame}
\vspace{-0.23in}
\begin{weirdFrame}{$\Enc_{\sk}(\bx)$}
Parse $(\pi,\bk,\sk') \leftarrow \sk.$

For each $i= 1,\dots,B:$
\begin{enumerate}
    \item Let $\bw_i = x_{ia + 1} \cdots x_{(i + 1)a}$.
    \item Generate $\br_i \unif \{0,1\}^{b}$.
    \item Let $\bz_i = f_{\bk}(i \circ \br_i)$.
    \item Let $\bw'_i = \RSE.\Enc_{\sk'}((\bw_i \oplus \bz_i) \circ \br_i)$.
\end{enumerate} 
Let $\by' = \bw'_1 \circ \dots \bw'_B$ and output $\by \leftarrow \pi(\by')$.
\end{weirdFrame}
\vspace{-0.23in}
\begin{weirdFrame}{$\Dec^{\Tilde{\by}}_\sk(L,R)$}
Parse $(\pi,\bk,\sk') \leftarrow \sk.$

Suppose $\bx[L,R]$ lies in $\bw_{i + 1} \circ \dots \circ \bw_{i + \ell}$ for some $i \in [B - \ell]$.
For each $j = i + 1, \dots, i + \ell$:
\begin{enumerate}
    \item Let $j_1,\dots,j_A$ be the indices of $\bw_j'$ in $\by'$.
    \item Let $\bw'_j = \widetilde{\by}_{\pi(j_1)}\circ\dots\circ\widetilde{\by}_{\pi(j_A)}$ be obtained by querying $\widetilde y$ at those $A$ indices.
    \item Compute $(\bd_{j,1} \circ \bd_{j,2}) \leftarrow \RSE.\Dec(\bw'_j)$ 
    \item Compute $\bw_j = \bd_{j,1} \oplus f_{\bk}(j \circ \bd_{j,2})$
\end{enumerate}
From $\bw_{i + 1},\dots,\bw_{i + \ell}$, output bits corr. to $\bx[L,R]$. 
\end{weirdFrame}
\end{construction}
    \begin{theorem}\label{thm:RSE_paLDC}
    Suppose $(\Gen_\RSE,\Enc_\RSE,\Dec_\RSE)$ is a $(A,a,\delta_\RSE)$-\RSE\ with rate $R_{\RSE} = a / A$. 
    Then, for any $q= \poly(\lambda)$, construction \ref{constr:RSE_paLDC} is a $(\alpha,\kappa,\delta,\eps,q)$-\paLDC\ for $\alpha = \frac{2 + o(1)}{R_\RSE}, \kappa = a, \delta < \delta_\RSE$, and $\eps = \negl(\lambda).$
    Furthermore, this code has rate $R = \theta(R_{\RSE})$.
\end{theorem}
    \renewcommand{\arraystretch}{1.5}
\begin{table}[htbp]
    \centering
\begin{tabular}{|l|l|}
    \hline
    {\em Hybrid's Codewords} & {\em Justification} \\ \hline
   $H_0: \{\pi (\RSE.\Enc_{\sk'}((\bw_j^i \oplus f_{\bk}(.)) \circ \br_j^i))_{j \in [B]}\}_{i \in [q]}$ & Original \\ \hline
   $H_1: \{\pi (\RSE.\Enc_{\sk'}((\bw_j^i \oplus \underline{\bR_j^i}) \circ \br_j^i ))_{j \in [B]}\}_{i \in [q]}$ & PRF Indist. \\ \hline
   $H_2: \{\pi (\RSE.\Enc_{\sk'}(\underline{\bR_j^{'i}}))_{j \in [B]}\}_{i \in [q]}$ & Same Dist. \\   \hline
   $H_3:\{\pi (\underline{\bR^{''i}_j})_{j \in [B]}\}_{i \in [q]}$ & \RSE\ Indist. \\ \hline
   $H_4: \{\underline{\pi^i} (\bR^{''i}_j)_{j \in [B]}\}_{i \in [q]}$ & Same Dist. \\ \hline
\end{tabular}

\,\

    \caption{Hybrid Summary for Theorem \ref{thm:RSE_paLDC}}
    \label{tab:paldcHybrid}
\end{table}

\begin{IEEEproof}     
    For any $L,R \in [k]$ such that $R - L + 1 \geq \kappa$, suppose $x[L,R]$ lie in blocks $w_{s+1},\dots,w_{s_{s + \ell}}$.
    Then, $\ell \leq \lfloor{\frac{R - L + 1}{a}}\rfloor + 1$.
    To recover each of these $w_j$ blocks from $\widetilde y$, the decoder accesses the corresponding encrypted block $w_j'$. 
    Thus, the decoder accesses $\ell A = \ell \times \frac{(a + b)}{R_\RSE}$ bits in total and we have that 
        \begin{align*}
        \alpha &\leq \frac{\ell(a + b)/(R - L + 1)}{R_\RSE} \\
               &\leq \frac{\left(\lfloor{\frac{R - L + 1}{a}}\rfloor + 1\right)( a + b)/(R - L + 1)}{R_\RSE} \\
               &\leq \frac{\left(\frac{1}{a} + \frac{1}{R - L + 1} \right)(a + b)}{R_\RSE}\\
               &\leq \frac{2 + 2\frac{b}{a}}{R_\RSE} \tag{$R - L + 1 \geq a$},
    \end{align*}
    where the term ${2b/a \in o(1)}$.
    
    We now upper bound $\Pr[\paldcgame(\calA,\lambda,\delta,\kappa,q) = 1]$ by upper bounding the the probability of the event $\mathtt{BAD} = \bigcup_{i\leq q, j \leq B} \mathtt{BAD}_{j}^i$ where $\mathtt{BAD}_j^i$ is the event that in round $i$  block $\bw'_j$ has more than $\delta_\RSE A$ errors. As long as the event $\mathtt{BAD}$ does not occur it is guaranteed that the local decoder will be successful in all rounds. 
    
    We proceed by defining a series of modified games (or Hybrids), where we argue that the incorrect decoding probability difference from the original game only differs negligibly. 

    We define the series of Hybrids $H_0$ to $H_4$ as follows:
    Denote round by superscript notation. Then,
    \begin{enumerate}
    \item $H_0$: The game is played as-is.
    \item $H_1$: Same as $H_0$, except that we update line 3 of the encoding algorithm to  $z_i \unif \{0,1\}^{A}$ i.e., we replace each pseudorandom string $f_{\bk}(j \circ \br_i)$ with a truly random string ${\bR_j \unif \{0,1\}^{A}}$ for each block $j \in [B]$.
    \item $H_2$: Same as $H_1$, except that we update line 4 of the encoding algorithm to $\bw_i'={\RSE.\Enc_{\sk'}(\bR_j)}$ where $\bR_j \unif \F^{a + \lg B}$ instead of $\bw_i'={\RSE.\Enc_{\sk'} (\bw_j \oplus \bR_j) \circ r_j^i}$.
    \item $H_3$: Same as $H_2$, but we update line 4 of the encoding algorithm to set $\bw_i\unif \F^A$ i.e., replacing $\RSE.\Enc_{\sk'}$ with a uniformly random string.
    \item $H_4$: Same as $H_3$, but in each round $i$ we sample a fresh permutation $\pi_i$ and output $\by = \pi_i(y')$. 
    \end{enumerate}
The indistinguishability of Hybrids $H_0$ and $H_1$ follow from PRF security and negligible collision probability.
First, disregarding possible collisions, the prf output $\bz_i^j = f_{\bk}(i \circ \br_i^j)$ is computationally indistinguishable from a random $\bR_j^i$.
Otherwise, a distinguisher for the prf can be constructed from the distinguisher of these Hybrids.
Next, there is added distinguishing probability when the same prf input is used across rounds i.e. when $j \circ \br_j^i = j \circ \br_j^{i'}$ for some $i \neq i' \in [q].$
Since the probability of a collision is at most $q^2/ 2^a,$ Hybrids $H_0$ and $H_1$ remain computationally indistinguishable.
Hybrids $H_1$ and $H_2$ are statistically indistinguishable because they are the same distribution.
Hybrids $H_2$ and $H_3$ are computationally indistinguishable by the security of \RSE. 
Recall that by the security property of \RSE, a computationally bound adversary cannot distinguish between polynomial-many random words and \RSE\ encodings with random inputs.
It follows that if Hybrids $H_2$ and $H_3$ were distinguishable, this would contradict the \RSE\ security property.
Lastly, Hybrids $H_3$ and $H_4$ are statistically indistinguishable since they form the same distribution. 
Justification for why each Hybrid is indistinguishable with respect to decoding error is summarized in the table \ref{tab:paldcHybrid}.

Thus, it suffices to upper bound the probability of the event $\mathtt{BAD}$ in Hybrid $H_4$ where a fresh random permutation $\pi_i$ is used in each round $i$. 

Since a new permutation $\pi^j$ with a uniformly random mask $\bR^{'' i}_j$ is used in every round $j$, an adversary's $\delta Ab$ errors are uniformly distributed in each block of size $A$.
    Thus, the number of errors for a given block $j$ is hyper-geometric$(AB, \delta AB, A)$, and by \cite{goldberg_public_2011,hush_concentration_2005}, we have
    \[\Pr[\mathtt{BAD}_{j}^i] < 2^{\frac{-2(((\delta_\RSE - \delta)A)^2 - 1)}{A + 1}}\]
    which is negligible with respect to $A = a / R_\RSE$ as long $\delta_\RSE > \delta$. 
    By applying a union bound over all $B$ blocks and $q$ rounds, we have that there is an error decoding any block is negligible.
\end{IEEEproof}
Theorem \ref{thm:RSE_paLDC} generically relies on a $(A,a,\delta_{\RSE})$-\RSE. For any $A=\omega(\log \lambda)$ we can construct an \RSE\ with $a=\Theta(A)$ and constant error tolerance $\delta_{\RSE} = \Theta(1)$. The probability that an attacker wins the \RSEGame\ is negligible in $\lambda$ and the constant rate $a/A = \Theta(1)$ is constant. Instantiating with this \RSE\ we obtain a $(\alpha,\kappa,\delta,\epsilon, q)$-paLDC with $\alpha=O(1)$, $\kappa = a =\omega(\log \lambda)$, $\delta= \Theta(1)$  and $\epsilon = \negl(\lambda)$.

\subsection{\aLDC s for Resource-bounded Channels}\label{subsect:Resource-boundedALDC}
Lastly, we present an \aLDC\ for resource-bounded channels with constant rate, constant amortized locality, and constant error-tolerance by applying the framework developed by Ameri et al. \cite{SCN:AmeBloBlo22} to eliminate the requirement that the encoder and decoder have a shared secret key. The framework of Ameri et al. \cite{SCN:AmeBloBlo22} using two building blocks: a secret key LDC and cryptographic puzzles. Intuitively, a cryptographic puzzle consists of two probabilistic polynomial time algorithms (PPT) $\mathtt{PuzzGen}$ and $\mathtt{PuzzSolve}$. $\mathtt{PuzzGen}(s)$ is a randomized algorithm that takes as input a string $s$ and outputs a puzzle $Z$ whose solution is $s$ i.e., $\mathtt{PuzzSolve}(Z)=s$. The security requirement is that for any adversary $A \in \mathcal{C}$ in a class $\mathcal{C}$ of resource bounded algorithms (e.g., bounded space, bounded computation depth, bounded computation constrained beyond PPT)  cannot solve the puzzle $Z$. In fact, we require that for any string $s_0$ and any resource bounded adversary $A \in \mathcal{C}$ the adversary $A$  cannot even distinguish between $(Z_0,s_0,s_1)$ and $(Z_1,s_0, s_1)$ where $s_i$ is a random string and $Z_i=\mathtt{PuzzGen}(s_i)$ is a randomly generated puzzle whose corresponding solution is $s_i$. 

At a high level the encoding algorithm $\Enc(x)$ works as follows: 1) pick a random string $r \in \{0,1\}^\lambda$ and generate a cryptographic puzzle $Z = \mathtt{PuzzGen}(r)$ whose solution is $r$. 2) Use a constant rate error correcting code to obtain an encoding $C_Z$ of this puzzle. 3) Use the random string $r$ to generate the cryptographic key $sk$ for a secret key \LDC\ (we will use the amortizeable secret key LDC  (\paLDC) from Theorem \ref{thm:RSE_paLDC}). 4) Use the secret key \LDC\ to encode the message and obtain $c_1=\Enc_{sk}(x)$. 5) Define $c_Z^1 = C_Z$ and $C_Z^{i+1}=C_Z \circ C_Z^{i}$ and find the smallest value $r$ such that $C_Z^r$ is at least as long as $c_1$. Set $c_2 = C_Z^r$. 6) Output the final codeword $C=c_1 \circ c_2$.  

Intuitively, if the channel is resource bounded then the channel cannot solve the puzzle $Z$ or extract any meaningful information about the solution $r$ or the secret key $sk$ derived from it. In contrast, the local decoding algorithm can extract several (noisy) copies of $C_Z$ by querying $c_2$ and decode these copies to extract $Z$ (most noisy copies of $C_Z$ in $c_2$ will still decode to $Z$). Then, the decoder, who runs in polynomial time but does not have the same resource constraints as the channel, can solve the puzzle $Z$ to obtain $r$ and then extract the secret key $sk$ using $r$. Finally, once the decoder has $sk$ it can run the (amortizeable) secret-key local decoder on $c_1$ to extract the message symbols that we want. 

If we instantiate this construction with a \paLDC, then the amortized locality is nearly the same. The local decoder needs to make $O(\lambda\ \mathtt{poly}(1/\epsilon))$ additional queries to $c_2$ to ensure that we recover the correct puzzle $Z$ with high probability e.g., at least $1-\epsilon/2$. However, these additional $O(\lambda\ \mathtt{poly}(1/\epsilon))$ queries can be amortized over the total number of symbols that are decoded. 
We can maintain constant amortized locality $\alpha$ as long as the amortization block size is larger than the key size. 
\begin{theorem}[Informal]
    Suppose the channel is resource-bounded and there exists a cryptographic puzzle.
    Suppose $\calC_P$ is a $(\alpha_p,\kappa_p,\delta_p,\eps_p,1)$-$\paLDC.$
    Then, under the framework of Ameri et al. \cite{SCN:AmeBloBlo22}, we can construct a ${(\alpha_p + o(1), \kappa_p, \delta,\eps)\text{-}\aLDC}$, where $\delta = O(\delta_p)$ and $\eps = O(\eps_p)$.
\end{theorem}

\section{Conclusion}\label{section:conclusion} 
We initiate the study of amortized \LDC s as a tool to overcome prior barriers and impossibility results. We show that it is possible to design an amortized \LDC\ with amortized locality $\alpha < 2$ --- overcoming an impossibility result of Katz and Trevisan for regular \LDC s. 
We also design a secret-key \LDC\ with constant rate, constant error tolerance, and constant amortized locality. 
Finally, under the natural assumption that the channel is resource bounded, we can use cryptographic puzzles to eliminate the requirement that the sender/ sender and obtain \LDC s with constant rate, constant error tolerance, and constant amortized locality. 

\bibliographystyle{IEEEtran}

\clearpage
\appendix
\subsection{Robust Secret Encyption from Code-based Cryptography}\label{app:RSE}
Code-based cryptography is a strong candidate for post-quantum cryptography due to the widely-believed intractability of code-theoretic problems related to decoding random linear codes\cite{berlekamp_inherent_1978,weger_survey_2024}. 
The most relevant to our work is the {\em average decoding problem assumption}, assuming that on average, it is hard for {\em any} adversary to distinguish between a codeword with a fixed number of errors and a uniformly random string.
\begin{assumption}[Average Decoding Hardness~\cite{EPRINT:DebRes22}]\label{ass:avgDecProb}
There exist functions $k = k(\lambda),$ and $n = n(\lambda)$ where $k = \poly(\lambda)$ and $n = \theta(k)$, and $n(\lambda) \geq k(\lambda)$ such that for any probabilistic polynomial time adversary $\calA$ playing the \ADPGame, there exists a negligible function $\eps$ such that for any $\lambda > 0,$
\[\left|\Pr[\ADPGame(\calA,n(\lambda),k(\lambda),\lambda) = 1] - \frac{1}{2}\right| < \eps(\lambda),\]
where the $\ADPGame$ is defined as,
\begin{weirdFrame}{$\ADPGame(\calA,n,k,\lambda)$}
Generate $\bx \unif \F^k, R \unif \F^{k \times n},$ $\be \leftarrow \F^n$ of hamming weight $\lambda,$ and $b \unif \{0,1\}$. 
If $b = 0$, send $\tilde{\bc} = \bx R + \be$ to the adversary $\calA$.
Otherwise, if $b = 1$, send $\bu \unif \F^n$.
The adversary outputs bit $b' \in \{0,1\}$, where if $b = b',$ the output of this experiment is $1.$ Otherwise, the output is $0.$
\end{weirdFrame}
\end{assumption}
Robert McEliece proposed the first code-based public-key cryptosystem in 1978 \cite{mceliece1978public}.
His construction was based on \textit{binary Goppa codes} and the assumption that such codes are sufficiently \textit{indistinguishable} from random codes. 
We state this assumption more formally by first defining Goppa codes and the Goppa code distinguishing (GD) game referenced from previous work.
    \begin{definition}[Binary Goppa Code \cite{berlekamp2003goppa}] 
    Let $g \in \F_{2^m}[x]$ be a polynomial of degree $\deg g \leq \lambda$ and let $L = \{\alpha_1,\dots,\alpha_n : g(\alpha_i) \neq 0\} \subset \F_{2^m}$ be a set of non-zero evaluation points over $g.$
    Then, a (binary) Goppa code $\calC$ with rate $\frac{n - m\lambda}{n}$ and error tolerance $\lambda$ is defined as
    \[\calC = \left\{\bc = (c_1,\dots,c_n) \in \F^n: \sum_{i = 1}^n \frac{c_i}{z - \alpha_i} \equiv 0 \bmod g(z)\right\}\]
\end{definition}

\begin{definition}[Goppa Code Distinguishing (GD) Game \cite{faugere_distinguisher_2013}]
\,\
\begin{weirdFrame}{$\gdGame(\calA,\lambda,n,m)$}
The challenger and adversary $\calA$ interact as follows:
\begin{enumerate}
\item Challenger chooses at random $b \unif \{0,1\}$. 
If $b = 0,$ then $G$ is set to be the generator of a random linear $[n,k]$-code.
Otherwise, $G$ is set to be the generator of a random $[n,k]$-Goppa code.
\item $G$ is given to $\calA,$ who outputs $b' \in \{0,1\}$.
If $b = b',$ the output of this experiment is $1.$
Otherwise, the output is $0.$
\end{enumerate}
\end{weirdFrame}
\end{definition}
The claim that Goppa code generator matrices have no structure discernible from a random matrix has withstood decades of cryptanalysis and attacks which have yet to disprove this assumption for general parameters~\cite{PQCRYPTO:BerLanPet08,leon_probabilistic_1988,EC:LeeBri88,faugere_distinguisher_2013}.
\begin{assumption}[Goppa-Random Indistinguishability]\label{ass:GoppaRandom}
There exist functions $m = m (\lambda), n = n(\lambda)$ and $k = n(\lambda) - \lambda m(\lambda)$ where $k = \poly(\lambda)$ and $n = \theta(k)$, and $n(\lambda) \geq k(\lambda)$ such that for
for any probabilistic polynomial time adversary $\calA,$ there exists a negligible function $\eps$ such that
\[\left|\Pr [\gdGame(\calA,\lambda,n,m) = 1] - \frac{1}{2}\right| < \eps(\lambda,n,m). \]
\end{assumption}

The security of the McEliece cryptosystem is contingent on Assumptions \ref{ass:avgDecProb} and \ref{ass:GoppaRandom}.
The cryptosystem will encode messages as codewords of a scrambled secret Goppa code with added noise, where decryption is straightforwardly undoing the scrambling and recover the message via the Goppa code decoder. 
More specifically, the secret key in the McEliece Cryptosystem is $\sk = (S,G,P),$ where $G$ is the generator of a chosen Goppa code, $P$ is a column permutation of $G$, and $S$ is an invertible message transformation. 
The public key is set to be the scrambled generator $\pk = SGP.$
The encryption scheme on input message $\bm,$ applies the scrambled encoding $(\bm SGP)$ and adds an error vector $\be$ with weight proportional to the security parameter $\lambda.$
The decryption scheme on input codeword with errors $\tilde{\bc} = (\bm SGP) + \be$, inverts the column permutation $\tilde{\bc} P^{-1} = \bm SG + \be P^{-1}$ and decodes the resulting word with the corresponding Goppa code decoder.

We construct a \RSE\ using a similar error-obfuscation coding scheme.
Our \RSE\ construction can be interpreted as a conversion of the McEliece system to the secret/symmetric key setting with additional consideration in the amount of errors added to support the desired \RSE\ error-tolerance. 
First, in the secret key setting, message scrambling matrix $S$ and the permutation matrix $P$ may be omitted.
While these matrices were used to disassociate the public Goppa code used in encoding and the secret Goppa code used in decoding, this is unnecessary when the encoder and decoder share the same secret key. 
Second, the \RSE\ robustness property requires that encodings can tolerate a $\delta$ fraction of errors.
In our code-based construction, this is naturally achievable by relaxing the weight of the added error vector $\be$ by a $\delta$ fraction. 

\begin{construction}\label{constr:Rse}
\begin{weirdFrame}{$\Gen(1^\lambda)$}
   Set $\sk \leftarrow G$, where $G \in \F^{k \times n}$ be the generator of a randomly chosen $[n,k]$ Goppa Code $(\Enc_{\msf{Gop}},\Dec_{\msf{Gop}})$ tolerant to $\lambda + \delta n$ errors.

\end{weirdFrame}

\begin{weirdFrame}{$\Enc_\sk(\bm)$}
    \begin{enumerate}
        \item Let $\bz \unif \F^n,$ where $wt(\bz) = \lambda$,
        \item Output $\bY \leftarrow \bm G + \bz.$ 
    \end{enumerate}
\end{weirdFrame}
\begin{weirdFrame}{$\Dec_\sk(\widetilde{Y})$}
    \begin{enumerate}
        \item From $(G,P) \leftarrow \sk$, compute $\inv P$.
        \item Compute $\widetilde X = \widetilde Y \inv P = (m\widetilde{GP} + z) \inv P = m \widetilde{SG} + \hat z,$ where $\hat z = z \inv P$. 
        Note that $wt(z \inv P) = \lambda + \delta n.$
        \item Output $\Dec_{G}(\widetilde X)$, where $\Dec_{G}$ is the decoding scheme (e.g Patterson's algorithm) for the Goppa code corresponding to $G$.
    \end{enumerate}
\end{weirdFrame}
\end{construction}
Note that the decoding of a Goppa code, in particular Patterson's algorithm, can be done in polynomial time with the extended Euclidean algorithm.
\begin{theorem}
Given Assumptions \ref{ass:avgDecProb} and \ref{ass:GoppaRandom}, Construction $\ref{constr:Rse}$ is a $(n,k,\delta)$-\RSE\ for any $n \geq k > 0$ and constant $\delta > 0$.
\end{theorem}
\begin{IEEEproof}
First, the robustness property is achieved by our encoding algorithm outputting a codeword $Y$ with $\lambda$ from a Goppa code that is tolerant to $\lambda + \delta n$ errors. 
Since the channel can induce at most $\delta n$ errors, the robustness property follows dierectly from the error tolerance of the Goppa code.

Next, we show that for any $q \in \poly(\lambda)$, all probabilistic polynomial time adversaries $\calA,$
\[\left| \Pr[\RSEGame(A,\lambda,q) = 1] - \frac 1 2\right| < \eps(\lambda).\]
We proceed by a hybrid argument over the probability of adversary $\calA$ winning the $\RSEGame$. 
We define the series of hybrids $H_0$ to $H_2$ as follows:
\begin{enumerate}
    \item $H_0$: The game is played as-is.
    \item $H_1$: Same as $H_0$, except in the key generation algorithm, replace the Goppa generator $G$ to a random code generator $R \unif \F^{k \times n}$.
    \item $H_2$: Same as $H_1$, except in encoding algorithm, replace the output in line $2$ with a random word $u \unif \{0,1\}^n.$
\end{enumerate}
Hybrid $H_0$ is indistinguishable to hybrid $H_1$ by Goppa-random indistinguishability in Assumption \ref{ass:GoppaRandom}.
More specifically, if hybrids $H_0$ and $H_1$ can be distinguished, then we can build a distinguisher for the \gdGame\ by applying the \gdGame\ challenge code (either a Goppa code or random code) to form challenge codewords for the \RSEGame.
Note that we can claim indistinguishability across all $q$ rounds immediately since the challenge messages are generated independently per round.
Next, hybrid $H_1$ is indistinguishable to $H_2$ by average decoding hardness, which follows immediately from the encoding algorithm in hybrid $H_1$, outputting $\bx R + \be$, where $wt(\be) = \lambda.$

In Hybrid $H_2$, the $b = 0$ and $b = 1$ cases are perfectly indistinguishable, and by our hybrid argument, the advantage of Hybrid $H_0$ is at most the advantage in Hybrid $H_2$ plus the negligibale advantage of the Hybrid in between.
Thus, the construction satisfies \RSE\ indistinguishability.
\end{IEEEproof}
\subsection{Detailing the One Time \paLDC}\label{app:oneTime}
\begin{construction}\label{constr:paLDCV1} 
Suppose that code $\calC = (\Enc_{\calC},\Dec_{\calC})$ is an $(A,a)$-code over an alphabet $\Sigma$ of size $q$. 
Let $c = \log q$.
\begin{weirdFrame}{$\Gen(1^\lambda) $}
    Output $\sk \leftarrow (\br,\pi) \in \{0,1\}^{cAB + |\pi| }$    

\end{weirdFrame}

\begin{weirdFrame}{$\Enc_{\sk}(\bx)$}
\begin{enumerate}
    \item Parse $(\br,\pi) \leftarrow \sk$.
    \item \textbf{Blocking.} Let $\bx = \bw_1 \circ \bw_2 \circ \dots \circ \bw_{B}$ where each $w_s \in \Sigma^a$ for $s = 1,\dots, B$. 
    \item \textbf{Block Encoding.} Encode each block $s$ as $\bw_s' = \Enc_{\calC} (\bw_s)$, let $\bx' = \bw_1'\circ\dots\circ \bw_B'$.
    \item \textbf{Permute and Mask.} Output $\pi(\bx') \oplus \br$.
\end{enumerate}
\end{weirdFrame}

\begin{weirdFrame}{$\Dec_{\sk}^{\Tilde{\by}}(L,R)$}
Parse $(\br,\pi) \leftarrow \sk$.
Let the interval $[L,R]$ of x bits lie in blocks $\bw_{s},\bw_{s+1},\dots,\bw_{s_{s + v}}.$ 
That is, for all $i \in [L,R]$, there exists $u \in [v]$ such that $x_i$ is a bit in $w_{s + u}$.
For each $j = s,\dots,s + v:$
\begin{enumerate}
    \item \textbf{Unmask.} Let $j_1,\dots,j_\ell,$ where $\ell = cA,$ be the indices of the bits corresponding to $\bw'_j$. 
    Compute $\Tilde{\bw}'_j = (y_{\pi(j_h)} \oplus r_{\pi(j_h)})_{h \in [\ell]}$.
    \item \textbf{Decode.} Apply $\Dec_{\calC}(\Tilde{\bw}'_j)$ to obtain $\Tilde{\bw_j}$. 
\end{enumerate}
From $\Tilde{\bw}'_{s_1},\dots,\Tilde{\bw}'_{s_v}$ output bits corresponding to interval $[L,R]$. 
\end{weirdFrame}
\end{construction}
\begin{theorem}\label{thm:paLDCV1}
    Suppose $\calC$ has rate $R_{\calC}$ and error tolerance $\delta_{\calC}$ and $a \in \omega(\log \lambda).$ 
    Then, Construction \ref{constr:paLDCV1} is a $\left(2/R,a,\delta,1\right)$-\paLDC\ when $\delta_{\calC} > \delta$.
\end{theorem}
\begin{IEEEproof}
    For any $L,R \geq 0$ such that $R - L + 1 \geq \kappa$, $x[L,R]$ lie in blocks $w_{s},w_{s+1},\dots,w_{s_{s + v}}$, where $v \leq \lfloor{\frac{R - L + 1}{ca}}\rfloor + 1$.
    We will show that our decoding process queries at most $2(R - L + 1) / R$ bits of the codeword i.e we show $\alpha \leq 2/R$.
    
    To recover these blocks, the decoder accesses $cAv$ bits and so,
    \begin{align*}
        \alpha &\leq \frac{cAv}{R - L + 1} \\
               &\leq \frac{cA\left(\lfloor{\frac{R - L + 1}{ca}}\rfloor + 1\right)}{R - L + 1} \\
               &\leq \frac{1}{R} + \frac{cA}{R - L + 1} \\
               &\leq \frac{1}{R} + \frac{cA}{ca}  \tag{$ca \leq R - L + 1 $}\\
               &= \frac{2}{R}.
    \end{align*}
    We show that the probability of an incorrect decoding $\Pr[\paldcgame(\calA,\lambda,\delta,\kappa,1) = 1]$ is negligible by upper bounding the the probability of the event $\mathtt{BAD} = \bigcup_{j \leq B} \mathtt{BAD}_{j}$ where $\mathtt{BAD}_j$ is the event that block $\bw'_j$ has more than $\delta_{\calC} A$ errors. 
    As long as the event $\mathtt{BAD}$ does not occur it is guaranteed that the local decoder will be successful in all rounds. 
    By Lipton's theorem\cite{STACS:Lipton94}, since our encoder applies a random mask and permutation, errors are added in a uniformly random manner with probability $\delta.$
    Thus, the number of errors for any given block $j$ is hypergeometric$(AB,\delta A B,A)$ and by \cite{goldberg_public_2011,hush_concentration_2005} we have 
    \[\Pr[\mathtt{BAD}_{j}] < 2^{\frac{-2(((\delta_{\calC} - \delta)cA)^2 - 1)}{cA + 1}},\]
    which is negligible for $\delta_{\calC} < \delta.$
    By a union bound over the total number of blocks, which is bounded by $\poly(\lambda)$, the probability, that any block fails its decoding is negligible. 
\end{IEEEproof}

\subsection{Resource-bounded \paLDC\ Construction}
Lastly, we present an \aLDC\ for resource-bounded channels with constant rate, constant amortized locality, and constant error-tolerance by applying the framework developed by Ameri et al. \cite{SCN:AmeBloBlo22} to eliminate the requirement that the encoder and decoder have a shared secret key. The framework of Ameri et al. \cite{SCN:AmeBloBlo22} using three building blocks: a secret key LDC, an $\LDCStar,$ and cryptographic puzzles. 

First, we formally introduce \LDCStar s, a variation on \LDC s in which the entire message is recovered while making few queries to the codeword (possibly with errors).
\begin{definition}[\LDCStar \cite{ITC:BloKulZho20}]\label{def:ldcStar}
    A $(n,k)$-code $\calC = (\Enc,\Dec)$ is an $(\ell,\delta,\eps)$-\LDCStar\ if for all $\bx \in \Sigma^k$, $\Dec^{\tilde{y}}$ with query access to word $\tilde\by$ $\delta$-close to $\Enc(\bx)$, 
    \[\Pr[\Dec^{\tilde{y}} = \bx] \geq (1 - \eps),\]
    where $\Dec^{\tilde{y}}$ makes at most $\ell$ queries to $\tilde{\by}$.
\end{definition}
Blocki et al. crucially show that an \LDCStar\ can be constructed for locality $\ell = O(k\poly(1/\eps))$ for arbitrarily large $n$ with constant decoding error.
Next, we formally defining cryptographic puzzles $\eps$-hard for algorithms $\R.$
\begin{definition}[\Puz \cite{SCN:AmeBloBlo22}]\label{def:puz}
    A puzzle $\Puz = (\Gen,\Sol)$ is a $(\R,\eps)$-hard puzzle for algorithm class $\R$ if there exists a polynomial $t'$ such that for all polynomials $t > t'$ and every algorithm $\calA \in \R,$ there exists $\lambda_0$ such that for al $\lambda > \lambda_0$ and every $s_0,s_1 \in \{0,1\}^\lambda$, we have 
    \[\left| \Pr \calA(Z_{b},Z_{1 - b},s_0,s_1) = b] - \frac{1}{2}\right| \leq \eps(\lambda),\]
    where the probability is taken over $b \unif \{0,1\}$ and $Z_i \leftarrow \Gen(1^\lambda,t(\lambda),s_i)$ for $i \in \{0,1\}.$
\end{definition}
Intuitively, a cryptographic puzzle consists of two algorithms $\Puz.\Gen$ and $\Puz.\Sol$. $\Puz.\Gen(s)$ is a randomized algorithm that takes as input a string $s$ and outputs a puzzle $Z$ whose solution is $s$ i.e., $\Puz.\Gen(Z)=s$. The security requirement is that for any adversary $A \in \R$ is a class $\R$ of resource bounded algorithms (e.g., bounded space, bounded computation depth, bounded computation) cannot solve the puzzle $Z$. In fact, we require that for any string $s_0$ and any resource bounded adversary $A \in \R$ the adversary $A$  cannot even distinguish between $(Z_0,s_0,s_1)$ and $(Z_1,s_0, s_1)$ where $s_i$ is a random string and $Z_i=\Puz.\Gen(s_i)$ is a randomly generated puzzle whose corresponding solution is $s_i$. 

We now modify the aforementioned compiler of Ameri et al. to take in a \paLDC\ instead of an \aLDC\ and to output a \aLDC\ instead of a \LDC\ for a resource-bounded channel. 
Additionally, we relax the definition of an $\aLDC$ to take in a consecutive range, like in $\paLDC$.
This follows naturally from the use of a \paLDC\ to instantiate our construction. 
Whether resource-bounded \aLDC s exist for non-consecutive queries is left as an open question.
Note that the codes defined are over choice of $\lambda$ values, where the message length is taken to be any polynomial $k \in \poly(\lambda).$ 
\begin{construction}\label{constr:resourcePaLDC}
Let $\paLDC.(\Gen,\Enc,\Dec)$ be a \paLDC\ that is a $(\nP,\kP)$-code, let $\LDCStar.(\Enc,\Dec)$ be a \LDCStar that is a $(\nStar,\kStar)$-code, and let $\Puz.(\Gen,\Sol)$ be a  $(\R,\eps)$-hard puzzle.
Let $t'$ be the polynomial guaranteed by Definition \ref{def:puz}.
Then, for any fixed $\lambda \in \N,$ we construct an \aLDC\ $(\Enc,\Dec)$ as the following algorithms.
    \begin{weirdFrame}{$\Enc(\bx)$}
        \begin{enumerate}
            \item Sample random seed $\bs \unif \{0,1\}^{\kP}$.
            \item Choose polynomial $t > t'$ and compute $\bZ \leftarrow \Puz.\Gen(1^\lambda,t(\lambda),s)$ where $\bZ \in \{0,1\}^{\kStar}.$
            \item Set $\YStar \leftarrow \LDCStar.\Enc(\bZ)$.
            \item Set $\sk \leftarrow \paLDC.\Gen(1^\lambda;\bs)$ i.e. we explicitly instantiate $\paLDC.\Gen$ with random seed $\bs$.
            \item Set $\YP \leftarrow \paLDC.\Enc_{\sk}(\bx)$
            \item Output $\YStar \circ \YP$.
        \end{enumerate}
    \end{weirdFrame}   

        \begin{weirdFrame}{$\Dec^{\widetilde{\YP} \circ \widetilde{\YStar}}(L,R)$}
        \begin{enumerate}
            \item Decode $\bZ \leftarrow \LDCStar.\Dec^{\widetilde{\YStar}}$.
            \item Compute $\bs \leftarrow \Puz.\Sol(\bZ)$
            \item Compute $\sk \leftarrow \paLDC.\Gen(1^\lambda;\bs)$
            \item Output $\paLDC.\Dec^{\widetilde{\YP}}_{\sk}(L,R)$
        \end{enumerate}
    \end{weirdFrame}    
\end{construction}
\begin{theorem}
    Suppose Construction \ref{constr:resourcePaLDC} is instantiated with an $(\alphaP,\kappaP,\deltaP,\epsP,q)$-\paLDC, $(\ellStar,\deltaStar,\epsStar)$-\LDCStar, and a $(\R,\epsPuz)$-hard puzzle \Puz. Then, Construction \ref{constr:resourcePaLDC} is a $(n,k)$-code that is a $\left(\alpha, \kappa, \delta,\eps \right)$-\aLDC\ with 
    \begin{align*}
        n &= \nP + \nStar,\\
        k &= \kP, \\
        \alpha &= \alpha_p + \frac{\ell_*}{\kappaP}, \\
        \kappa &= \kappaP,\\
        \delta &= (1/n) \times \min\{\deltaStar \nStar, \deltaP \nP\}, \\
        \eps &= ((1 - \epsStar) \epsP + 2 \epsPuz) /\epsStar,
    \end{align*}
    when the adversarial channel $\calA \in \R.$
\end{theorem}
\begin{IEEEproof}
    The decoder given $L,R \geq 0$ such that $R - L + 1 \geq \kappaP,$ queries the word $\tilde{\YStar} \circ \tilde{\YP}$ in two steps. 
    First, it performs queries for $\LDCStar.\Dec^{\tilde{\YStar}}$ to recover the puzzle $\bZ$, and after solving the puzzle to generate secret key $\sk,$ it performs queries for $\paLDC.\Dec_{\sk}^{\tilde{\YP}}(L,R)$. 
    The total number of queries is at most $\ellStar + (R - L + 1) \alphaP$ so $\alpha$ can be derived as
    \[\alpha (R - L + 1) \leq \ellStar + (R - L + 1)\alphaP \implies \alpha \leq \alphaP + \ellStar / \kappaP.\]
    Next, the error tolerance $\delta$ can be derived as the weighted minimum error-tolerance between the \LDCStar\ and \paLDC\ codewords, $\delta = (1/n) \times \min\{\deltaStar \nStar, \deltaP \nP\}.$
    A channel can choose either $\YP$ or $\YStar$ to place all $\delta n$ errors, so the fraction of errors tolerated is the minimum of the error tolerances of the chosen \LDCStar \ and \paLDC. 
    
    The remainder of the proof of security and decoding error follows the proof of Theorem $6.8$ of \cite{SCN:AmeBloBlo22} with minor changes for notational differences.
    We repeat it for the sake of completeness.
    Suppose there exists an algorithm/adversarial channel $\calA \in \R$ that, given the hard puzzle \Puz, can cause a decoding error for some chosen range $[L,R]$ with some non-negligible probability $\eps(\lambda)$ for some $\lambda \in \N.$
    Then, we can construct an adversary $\calB \in \R$ breaking security of the $(\R,\epsPuz)$-hard puzzle by a two-phase hybrid argument.
    In the first phase, we define two encoders $\Enc_0$ and $\Enc_1$, where $\Enc_0$ is equal to the original encoder in Construction \ref{constr:resourcePaLDC} and $\Enc_1$ is defined to take in the secret key $\sk$ (rather than generating it as in $\Enc_0$) as follows.
    \begin{weirdFrame}{$\Enc_1(\bx,\sk)$}
                \begin{enumerate}
            \item Sample random seed $\bs \unif \{0,1\}^{\kP}$.
            \item Choose polynomial $t > t'$ and compute $\bZ \leftarrow \Puz.\Gen(1^\lambda,t(\lambda),s)$ where $\bZ \in \{0,1\}^{\kStar}.$
            \item Set $\YStar \leftarrow \LDCStar.\Enc(\bZ)$.
            \item Set $\YP \leftarrow \paLDC.\Enc_{\sk}(\bx)$
            \item Output $\YStar \circ \YP$.
        \end{enumerate}
    \end{weirdFrame}
    We proceed with constructing the two-phase hybrid distinguisher $(\calD_1,\calD_2)$.
    First, $\calD_1$ is given as input message $\bx$ and access to encoders $\Enc_0$ and $\Enc_1.$
    \begin{weirdFrame}{$\calD_1(\bx)$}
        \begin{enumerate}
            \item Compute $b \unif \{0,1\}$ and $\sk_1 \leftarrow \paLDC.\Gen(1^\lambda)$.
            \item If $b = 0$, output $\bY_b \leftarrow \Enc_0(\bx)$. 
            Otherwise, if $b = 1,$ output $\bY_b \leftarrow \Enc_1(\bx,\sk)$.
        \end{enumerate}
    \end{weirdFrame}
    Denote the secret key used in $\Enc_b$ as $\sk_b$, where $b \in \{0,1\}$. 
    $\sk_b$ is either generated by the distinguisher (when $b = 1)$ or the encoding algorithm $\Enc_0$ (when $b = 0)$.
    The output of $\calD_1(\bx)$ is given to the adversarial channel $\calA$, who outputs  $\bY_b' = \bY'_{\msf{P},b} \circ \YStar'$.
    Then, in phase two, $\calD_2$ is given as input the message $\bx$, the secret key $\sk_b$, and the corrupt \paLDC\ codeword $\bY'_{\msf{P},b},$ where $b$ is the bit computed by $\calD_1(\bx).$
    \begin{weirdFrame}{$\calD_2(\bx,\sk_b,\bY'_{\msf{P},b})$}
        \begin{enumerate}
            \item Sample $L,R \unif [k]$ such that $L \leq R$ and $R - L + 1 \geq \kappa.$
            \item Compute $\bx_i' \leftarrow \paLDC.\Dec^{\bY'_{\msf{P},b}}_{\sk_b}(L,R)$.
            \item Output $b' = 0$ if $\bx_i \neq \bx_i'$, otherwise output $b' = 1.$
        \end{enumerate}
    \end{weirdFrame}
    Now, we give the two-phase distinguisher which breaks the $(\R,\epsPuz)$-hard puzzle. 
    For puzzle solutions $s_0,s_1$ we construct an adversary $\calB \in \R$ which distinguishes $(\bZ_b,\bZ_{1 - b},s_0,s_1)$ with probability at least $\eps'$ for $b \unif \{0,1\}$.
    Fix a message $\bx$ and $\lambda \in \N.$
    \begin{weirdFrame}{$\calB(\bZ_b,\bZ_{1 -b},s_0,s_1)$}
        \begin{enumerate}
            \item  \textbf{Encoding.}
        \begin{enumerate}
            \item Generate $\sk \leftarrow \paLDC.\Gen(1^\lambda;s_0)$.
            \item Set $\YStar \leftarrow \LDCStar.\Enc(\bZ_b)$.
            \item Set $\YP \leftarrow \paLDC.\Enc_{\sk}(\bx)$.
            \item Set $\bY = \YStar \circ \YP$.
        \end{enumerate}
        \item Give $(\bx,\bY,\delta,\eps,k,n)$ to $\calA$ to obtain $\bY'$. 
        Set $\YP'$ as the substring corresponding to the corruption of $\YP$ above.
        \item Compute $x'_L,\dots, x'_R \leftarrow \paLDC.\Dec_{\sk}^{\YP'}$ and if there exists $x'_i \neq x_i$ for some $i \in [L,R],$ output $b' = 0$. 
        Else output $b' = 1.$
        \end{enumerate}
    \end{weirdFrame}
    First, by assumption, the adversary $\calB \in \R$ since each of its subroutines $\calA, \paLDC.\Enc, \paLDC.\Dec,\LDCStar \in \R.$
    We claim that $\calB$ distinguishes $(\bZ_b,\bZ_{1 -b},s_0,s_1)$ with noticeable advantage. 
    First, observe that $\sk$ is always generated by $\paLDC.\Gen(1^\lambda,s_0).$
    Next, for $b = 1$, $\YStar$ encodes puzzle $\bZ_1$ and the secret key $\sk$ is unrelated to the solution $s_1$ of puzzle $\bZ_1.$
    Since in this case, $\sk$ and $\YStar$ are uncorrelated, $\calA$ causes a decoding error with probability $\epsP$ where $\bY'$ is at most $\delta n$ distance away from $\bY.$

    In the case that $b = 0,$ puzzle $\bZ_0$ is encoded as $\YStar$ with solution $s_0$ that is used to generate $\sk.$ 
    Thus, in this case, the probability that the decoder outputs at least one incorrect $x_i$ is at least the advantage of $\calA$, $\eps.$

    $\calB$ outputs $b' = b$ when $b = 0$ with probability $\eps \epsStar (1/k)$ by the argument above. 
    For $b = 1,$ this probability is at least $1 - \epsP \epsStar$ since the probability that $b' = 0$ in this case is at most $\epsP \epsStar.$ 
    Thus, we have that 
    \[\Pr[\calB(\bZ_b,\bZ_{1 -b},s_0,s_1) = b] - \frac{1}{2} \geq \frac{1}{2}\left(\eps \epsStar (1/k) - \epsP (1 - \epsStar)\right),\]
    which is non-negligible, contradicting the security of the puzzle \Puz.
\end{IEEEproof}
Explicitly, we can achieve constant decoding error $\eps$, constant rate $R$, constant amortized locality $\alpha$, and constant error-tolerance $\delta$ by using the (one-time) \paLDC\ construction from Theorem \ref{thm:paLDCV1} and an \LDCStar\ with $\nStar \sim \nP.$ 
Note that $\nStar = \theta(\nP)$ is necessary for simultaneous optimally constant parameters.  
We need $\nStar = \Omega(\nP)$ for constant error tolerance, and, at the same time, the rate $R = \kP / (\nP + \nStar)$ is constant for $\nStar = O(\nP)$. 
\end{document}